\input harvmac
\noblackbox

\font\ticp=cmcsc10
 
\def\Title#1#2{\rightline{#1}\ifx\answ\bigans\nopagenumbers\pageno0\vskip1in
\else\pageno1\vskip.8in\fi \centerline{\titlefont #2}\vskip .5in}

\font\ticp=cmcsc10
\font\ttsmall=cmtt10 at 8pt

\input epsf
\ifx\epsfbox\UnDeFiNeD\message{(NO epsf.tex, FIGURES WILL BE
IGNORED)}
\def\figin#1{\vskip2in}
\else\message{(FIGURES WILL BE INCLUDED)}\def\figin#1{#1}\fi
\def\ifig#1#2#3{\xdef#1{Fig.~\the\figno}
\goodbreak\topinsert\figin{\centerline{#3}}%
\smallskip\centerline{\vbox{\baselineskip12pt
\advance\hsize by -1truein\noindent{\bf Fig.~\the\figno:} #2}}
\bigskip\endinsert\global\advance\figno by1}

%
%
\def\p{\partial}
\def\[{\left [}
\def\]{\right ]}
\def\({\left (}
\def\){\right )}
\def\BH{black hole}
\def\GT{gauge theory}
\def\AfSf{$AdS_5 \times S^5$}
\def\AfSs{$AdS_4 \times S^7$}
\def\AsSf{$AdS_7 \times S^4$}
\def\g0{\buildrel \circ \over g}

\def\hl{{\hat l}}
\def\<{\langle}
\def\>{\rangle}
\def\O{{\cal O}}
\def\OZn{\< \O \, Z_l \>_n}


\lref\dkk{U. Danielsson, E. Keski-Vakkuri and M. Kruczenski, {\it
Spherically Collapsing Matter in AdS, Holography, and Shellons},
Nucl. Phys. B563 (1999) 279, hep-th/9905227.}
\lref\dkkk{U. Danielsson, E. Keski-Vakkuri and M. Kruczenski, {\it
Vacua, Propagators, and Holographic Probes in AdS/CFT},
JHEP 9901 (1999) 002, hep-th/9812007.}
\lref\simon{V. Balasubramanian and S. F. Ross,
{\it Holographic Particle Detection},
 Phys. Rev. D61 (2000) 044007, 
hep-th/9906226}
\lref\juan{J. Maldacena, 
{\it The Large N Limit of Superconformal Field Theories and Supergravity},
Adv. Theor. Math. Phys. 2 (1998) 231, hep-th/9711200.}
\lref\magoo{O. Aharony, S.S. Gubser, J. Maldacena, H. Ooguri, Y. Oz,
{\it Large N Field Theories, String Theory and Gravity},
Phys. Rept. 323 (2000) 183, hep-th/9905111.}
\lref\grla{R. Gregory and R. Laflamme, 
{\it Black Strings and p-Branes are Unstable},
Phys. Rev. Lett. 70 (1993) 2837, 
hep-th/9301052; 
{\it The Instability of Charged Black Strings and p-Branes},
Nucl. Phys. B428 (1994) 399, 
hep-th/9404071.} 
\lref\hoit{G. Horowitz and N. Itzhaki, 
{\it Black Holes, Shock Waves, and Causality in the AdS/CFT Correspondence},
JHEP 9902 (1999) 010, hep-th/9901012}
\lref\gary{G. Horowitz, 
{\it Comments on Black Holes in String Theory},
Class. Quant. Grav. 17 (2000) 1107, hep-th/9910082.}
\lref\peet{A. Peet and J. Polchinski, 
{\it UV/IR Relations in AdS Dynamics},
Phys. Rev. D59 (1999) 065011, 
hep-th/9809022.} 
\lref\suwi{L. Susskind and E. Witten, 
{\it The Holographic Bound in Anti-de Sitter Space},
hep-th/9805114.} 
\lref\gst{J. McGreevy, L. Susskind, and N. Toumbas, 
{\it Invasion of the Giant Gravitons from Anti-de Sitter Space},
JHEP 0006 (2000) 008,
hep-th/0003075.} 
\lref\gmt{M. Grisaru, R. Myers, and O. Tafjord,
{\it SUSY and Goliath},
hep-th/0008015.} 
\lref\hhi{A. Hashimoto, S. Hirano and N. Itzhaki,
{\it Large branes in AdS and their field theory dual},
hep-th/0008016.}
\lref\mast{J. Maldacena and A. Strominger, 
{\it AdS3 Black Holes and a Stringy Exclusion Principle},
JHEP 9812 (1998) 005,
hep-th/9804085.}
\lref\joe{J. Polchinski,
{\it S-Matrices from AdS Spacetime},
hep-th/9901076.}
\lref\suto{L. Susskind and N. Toumbas,
{\it Wilson Loops as Precursors},
Phys.Rev. D61 (2000) 044001, hep-th/9909013 .}
\lref\bisu{D. Bigatti and L. Susskind,
{\it TASI lectures on the Holographic Principle},
hep-th/0002044.}
\lref\pst{J. Polchinski, L. Susskind, and N. Toumbas,
{\it Negative Energy, Superluminosity and Holography},
Phys.Rev. D60 (1999) 084006, hep-th/990322.}
\lref\lmr{J. Louko, D. Marolf, and S. F. Ross,
{\it On geodesic propagators and black hole holography},
Phys.Rev. D62 (2000) 044041, hep-th/0002111.}
\lref\steve{S. B. Giddings,
{\it Flat-space scattering and bulk locality in the AdS/CFT correspondence},
Phys.Rev. D61 (2000) 106008, hep-th/9907129.}
\lref\lenny{L. Susskind,
{\it Holography in the Flat Space Limit},
hep-th/9901079.}
\lref\giro{S.  Giddings and S. Ross,
{\it D3-brane shells to black branes on the Coulomb branch},
Phys.Rev. D61 (2000) 024036,
hep-th/9907204.}

\baselineskip 16pt
\Title{\vbox{\baselineskip12pt
\line{\hfil \tt hep-th/0009051} }}
{\vbox{
{\centerline{CFT Description of Small Objects in AdS}}
}}
\centerline{\ticp Gary T. Horowitz and Veronika E. Hubeny\footnote{}{\ttsmall
gary@cosmic.physics.ucsb.edu, veronika@cosmic.physics.ucsb.edu}}
\bigskip
\centerline{\it Physics Department, University of California,
Santa Barbara, CA 93106, USA}
\bigskip
\centerline{\bf Abstract}

By the AdS/CFT correspondence, the expectation value of certain local
operators in
the CFT is given by the asymptotic value of  supergravity fields. We show
that  these local expectation values contain a 
remarkable amount of information about
small sources deep inside $AdS_p\times S^q$. 
In particular, they contain essentially all the 
multipole moments. More importantly, one can use them to determine the size of
a spherical source. This is not a small effect: The size appears in an 
exponentially large contribution to the expectation values.
This provides an easy way for the CFT to distinguish stars 
from black
holes with the same mass, or to distinguish different ``giant graviton"
configurations.

\bigskip

\Date{September, 2000}
\newsec{Introduction}

One of the main open problems in the AdS/CFT correspondence \refs{\juan,\magoo}
is the following:
What is the proper conformal field theory (CFT) description of  objects
in the bulk with size much less than the anti de Sitter (AdS) radius $R$?
There is a hope that since physics on scales much less than $R$ should be
similar to that in 
flat space, this description would be an important step toward
finding a holographic theory for asymptotically flat spacetimes.
(For some work in this direction, see \refs{\joe,\steve}.) 
We find evidence that this will not be 
the case. Special properties of the asymptotic
$AdS_p\times S^q$ boundary conditions appear to be crucial in the CFT 
description of small objects in the bulk spacetime.

By the UV/IR connection \refs{\suwi,\peet},
one might expect that the description of small objects 
will require highly nonlocal operators \refs{\lenny,\bisu}.
Surprisingly, this is not quite the case. Even though
the expectation value of local operators in the gauge theory are sensitive
only to the asymptotic value of supergravity fields, we will show that these
leading order 
asymptotic fields contain a remarkable amount of information about small
sources deep inside the spacetime.

In this respect, asymptotically $AdS_p\times S^q$
spacetimes are dramatically different
from the more familiar asymptotically flat spacetimes.
In the latter case, it is well known that
the multipole moments of a source can be read-off from the asymptotic
behavior of the field. But since higher multipole moments fall off faster,
the leading asymptotic field depends only 
on the monopole moment 
- the total
mass. In particular, there is no information about the size of the source.
For a small source in $AdS_p\times S^q$, one might have expected
that the situation would be very similar. After all, if the size $r_0$ of the
source is much less than the radius of curvature $R$, there is an
intermediate region $r_0\ll r\ll R$ in which the spacetime is approximately
flat and the information about the higher multipole moments will fall off
more quickly. Nevertheless, we will show that for $r\gg R$,
the situation is, in fact, very different. The $AdS_p\times S^q$ boundary
conditions effectively ``refocus" information about the source.
It turns out that one can recover
essentially all multipole moments of the source from
the {\it leading} asymptotic behavior of the field. More importantly,
one can recover the size of a spherical source from this asymptotic field.

We will focus mainly on the case of $AdS_5\times S^5$ boundary conditions,
which is described by an ${\cal N} =4$ super Yang-Mills theory, although
we will comment on other values of $p$ and $q$ where similar results 
are obtained. To illustrate the effect, we will consider the simplest
possible case of a massless scalar field $\Phi$ with source $s$
in $AdS_5\times S^5$. This can be viewed as the dilaton. Perturbations of 
the metric or other supergravity fields should behave similarly.

It may seem surprising that one can say anything about the size of a source
given the asymptotic value of the field. After all, given any static
solution to the wave equation $\Phi$
in a neighborhood of infinity, one can continue it
inside satisfying the field equation for a while. Then, at an arbitrary radius,
one can
take any smooth continuation of $\Phi$  to the origin
and define $s \propto \nabla^2 \Phi$.
The result is a source, $s$, which produces the
field $\Phi$.
So the size of the source is not in general
fixed by the asymptotic field. However,
in our case there is an additional constraint coming from symmetry. A small
source which is spherically symmetric in ten dimensions
will produce a field in $AdS_5
\times S^5$ which depends only on a radial distance $\chi$ in $S^5$ and a
radial distance $\rho$ in $AdS_5$. One could take this asymptotic field,
cut it off at an arbitrary distance, extend it to the origin in
any smooth way, and define a new source $s \propto \nabla^2 \Phi$. But this
source will in general be a function of both  $\rho$ and $\chi$. It will
not be  approximately spherically symmetric. The requirement that the
source be just a function of $r^2 \equiv \rho^2 + \chi^2$ places a strong
constraint on how the asymptotic field can be matched onto a source. In
particular, we will show that it fixes the size of the source.

The presence of the $S^5$ is crucial for obtaining this size information.
Since a small source of size $r_0 \ll R$ breaks the $SO(6)$ symmetry, there
will be nonzero expectation values of operators
involving $Z_l \equiv V_{j_1 \cdots j_l} X^{j_1} \cdots X^{j_l}$, where $V$
is a symmetric traceless tensor of rank $l$, and $X^j$ are the six scalars
in the
${\cal N}=4$ supersymmetric gauge theory.
Given that the field in the intermediate region
$r_0 <r<R$ is dominated by the monopole moment, one might have
expected that, for certain local operators $\O$,
$\<\O Z_l\>$ would be universal and determined only
by the mass of the source. However,
this is not the case. We will see that the behavior of
the expectation 
values $\<\O Z_l\>$ changes dramatically when $l\sim (R/r_0)^2$.\foot{This is
for spherical sources. For non-spherical sources, we will see that the
change occurs at $l\sim (R/r_0)$.}
They increase
like  $l^2$ for  $l\ll (R/r_0)^2$, but grow exponentially with $l$ for
$l\gg (R/r_0)^2$! 
Thus it is easy to extract information about the size of the source.

Since we expect metric perturbations (which couple to the stress energy
of all sources) to behave similarly to the scalar field $\Phi$,
there are several possible applications of this result. 
Trying to describe a small object in spacetime just from local operators in 
the gauge 
theory is analogous to astrophysicists trying to understand an exotic object
in a distant galaxy by observing the radiation emitted. While it is relatively
easy to get information about the mass of the object, it is usually difficult to
directly measure its size.  With $AdS_5\times S^5$ boundary conditions, this
difficulty is removed. So it is easy to determine that the source is a collapsed
object rather than, e.g., a star. Another possible application concerns
expanded brane configurations. It has recently been shown \refs{\gst,\gmt,\hhi}
that there are three
different BPS configurations with the same mass and angular momentum as
a graviton. In addition to the usual pointlike graviton, there is a three-brane
expanded the $AdS_5$ directions, and another
expanded in the $S^5$ directions. The
question was raised in \hhi\ as to how the CFT could distinguish these different
configurations. Although we will focus on static sources, and not the moving
branes required for these ``giant gravitons", the results described here point
toward a clear distinction between these three cases.

\newsec{Recovering the source from the leading asymptotic field}

The basic reason one can
obtain information about small sources in $AdS_5\times S^5$ from
the leading order field at infinity is the following. Unlike asymptotically
flat spacetimes,
higher multipole moments of a field in $AdS_5$ do {\it not} 
fall off faster
at infinity \simon. All modes of a field fall off at the same rate, so from
the asymptotic value of the field at infinity,
one can recover all the multipole moments of the source
in the $AdS_5$ directions. The $S^5$ dependence of the source
can be expanded in $S^5$ spherical harmonics (since they form a complete
basis of functions). For sources which are a product of a function on
$S^5$ and a function on $AdS_5$,
if we could recover all the coefficients of this
expansion, we could reconstruct not just the multipole moments but the entire
source function in the $S^5$ directions. This is not quite possible, but from
the asymptotic values of the corresponding Kaluza-Klein fields
in $AdS_5$, we show below that even for general source functions,
one can recover considerable information about the source in the
$S^5$ directions, including its size.

Since the Kaluza-Klein modes will play a crucial role, we
begin by deriving a simple property of solutions to 
massive wave equations in general static
spacetimes.
Consider a massive wave equation with source $s$
\eqn\Poiss{\nabla^2 \Phi -m^2 \Phi = -k s}
in a d-dimensional static spacetime with metric
$ds^2 = - f^2 dt^2 + g_{ij} dx^idx^j$. (The constant $k$ is fixed by the 
normalization of $\Phi$.)
Static solutions to the wave equation
satisfy\foot{This
can easily be seen by writing $\nabla^2 \Phi = g^{-1/2} \p_\mu(g^{1/2}
g^{\mu\nu} \p_\nu \Phi) $
where $g = f^2 \det{g_{ij}}$ is the determinant of the 
spacetime
metric.}
\eqn\laplace{D_i(fD^i \Phi)  -m^2 f \Phi = -k s f}
where $D_i$ is the covariant derivative on a $t=$ constant surface $\Sigma$.
The standard flat space identities relating integrals over the source to the
asymptotic value of the field can easily be generalized to curved space. 
Let $u$ be any nonsingular solution of the source-free equation 
\eqn\ueq{D_i(fD^i u) -m^2 f u =0.}
Then we can multiply \laplace\ by $u$ and integrate over the spatial 
surface $\Sigma$ to get 
\eqn\multipole{k \int_\Sigma  u s f =  
  \oint_S \Phi fn^i \p_i  u - \oint_S u f n^i \p_i \Phi}
where the integral on the right is over a sphere at infinity and
$n^i$ is its unit normal. This gives us a relation between an integral of
the source and the asymptotic value of the field $\Phi$.

We now consider some examples.
For a massless field in  flat spacetime, we can set 
$u_l =r^l Y_l(\Omega_{d-2})$
where $Y_l(\Omega_{d-2})$ is a spherical harmonic on $S^{d-2}$.
Since $f=1$,
the left hand side of \multipole\ is ($k$ times one component of) the $l^{th}$
multipole moment $M_l$. The
right hand side will be finite only if the part of the field  $\Phi$ with
angular dependence $ Y_l$ falls off like
$1/r^{d+l-3}$. Thus we recover the usual flat space result that higher
multipole moments of the source are encoded in higher order terms in the
asymptotic field.

Now consider $AdS_5$ with metric
$
ds^2  \, = - f^2(\rho) \, dt^2 + f(\rho)^{-2} \, d\rho^2
+ \rho^2 \, d\Omega_3^2$
with $ f^2(\rho) = {\rho^2 \over R^2} + 1$ where $R$ is the radius of curvature.
As noted in \simon, all static solutions to the massless wave equation go to a
constant plus $ O(1/\rho^4)$ {\it regardless of their angular dependence}.
(This will be reviewed below.) For normalizability, we require that the
constant is zero, so
\eqn\asymphi{\Phi(\rho,\Omega_3) \rightarrow {\phi(\Omega_3)\over \rho^4}}
Let $u_{\hat l}$ be the static solution that behaves like 
$\rho^{\hat l} Y_\hl(\Omega_3)$ near the origin. 
If the size of the source is much smaller than $R$, then $f \approx 1$ over
the extent of the source, and the left hand side of \multipole\
is still a multipole moment $M_\hl$.
Since $u_\hl$ now goes to a constant asymptotically, and $f\approx \rho/R$,
$n^i \p_i \approx (\rho/R) \p/\p \rho$ for large $\rho$,
we see that $\phi(\Omega_3)$
contains all the information about the multipole moments of the source.
More specifically, since $u_\hl$ will grow like $\rho^{\hat l}$
until $\rho \sim R$, and then
approach a constant asymptotically, we have $u_\hl \sim R^\hl$ for large
$\rho$. Plugging this into \multipole\ we see that the first term on the
right vanishes and we have
\eqn\estimate{ \int_{S^3} \phi Y_\hl \sim {M_\hl \over R^{\hl -2}}}
So for $R$ much larger than the size of the source $\rho_0$,
the modes of the asymptotic field with $\hl > 2$ are 
suppressed relative to the multipole moments, while for $\hl < 2$ they are
enhanced. (This can also be seen by dimensional analysis.) So even though
all the multipole moments can be recovered from $\phi(\Omega_3)$,
most of them make a very small contribution to the asymptotic field
when $\rho_0 \ll R$. 
This is consistent
with the fact that in the approximately flat region, $\rho_0\ll 
\rho \ll R$, the modes of the field decay faster with increasing $\hat l$.
However, the key
point is that all the multipole moments can in principle be recovered from
$\phi(\Omega_3)$.

We  now turn to the case of interest, $AdS_5\times S^5$. We will henceforth
set $R=1$ and measure all quantities in units of the AdS radius. The spacetime
metric is thus
\eqn\adsmetric{
ds^2  \, = - (\rho^2 +1) \, dt^2 +  {d\rho^2\over \rho^2 +1}
+ \rho^2 \, d\Omega_3^2
 +   \, d\Omega_5^2.}
We again assume that the 
size of the source $s$ is much smaller than the radius of curvature.
Since spacetime is essentially flat near the
source, we could take $u$ near the source
to be a spherical harmonic on $S^8$ times a
power of the radius. Then, the right hand side of \multipole\
would just be a (ten dimensional) multipole moment of the source. 
However, since the spacetime
metric is a product of $S^5$ and  $AdS_5$, it is much more convenient to 
expand all fields in spherical harmonics on $S^5$ and $S^3$.

 The standard mode decomposition of a static scalar field in \AfSf\ is
 \eqn\sepvar{
 \Phi(\rho,\Omega_3,\Omega_5) = \sum_{L, \hat L}
  {1 \over \rho^{3/2}} \, \psi_{L,\hat L}(\rho) \, Y_{\hat L}(\Omega_3)
  \, Y_L(\Omega_5)}
  where $Y_{\hat L}(\Omega_3)$ and $Y_L(\Omega_5)$ are the spherical
  harmonics on the $S^3$ and $S^5$, respectively, and $L,\hat L$ label the
  different modes, e.g., $L = (l, \{ m_i\})$. The spherical harmonics
  satisfy
\eqn\evalue{ \nabla^2_{\Omega_3} Y_{\hat L}(\Omega_3)
= - \hat{l} (\hat{l} + 2) \, Y_{\hat L}(\Omega_3) }
and 
\eqn\egvalue{\nabla^2_{\Omega_5} Y_L(\Omega_5)
= - l(l+4) \,  Y_L(\Omega_5)} 
The source can be similarly decomposed:
\eqn\source{
s(\rho,\Omega_3,\Omega_5) = \sum_{L, \hat L}
 {1 \over \rho^{3/2}} \, \sigma_{L, \hat L}(\rho) \, Y_{\hat L}(\Omega_3)
 \, Y_L(\Omega_5)}

 The radial equation for $\psi_{L,\hat L}(\rho)$ resulting from \Poiss\ 
 with $m=0$ then becomes
 \eqn\rad{
 \(\rho^2  + 1\) \, \psi_{L, \hat L}'' 
+ 2 \rho  \, \psi_{L, \hat L}' -
 V_{L, \hat L} \, \psi_{L, \hat L} = -k \sigma_{L, \hat L}}
 where each term has an implicit $\rho$ dependence and
\eqn\V{
V_{L, \hat L}(\rho) =
{3 \over 4\rho^2} + {15 \over 4 } 
 + {\hat{l} (\hat{l} + 2) \over \rho^2} + l(l+4) }
Asymptotically, i.e.\ as $\rho \to \infty$,
$V_{L,\hat L}(\rho) \to {15 \over 4 } + l(l+4) $
and $\sigma_{L,\hat L}(\rho)= 0$,
so that \rad\ simplifies to
\eqn\radas{
\rho^2 \, \psi_{L, \hat L}'' + 2 \rho  \, \psi_{L, \hat L}' -
\( {15 \over 4} +l(l+4) \) \psi_{L, \hat L} = 0}
This has the solution
\eqn\assol{
\psi_{L, \hat L}(\rho) \sim 
 \rho^{-\( l+ {5 \over 2} \)}
}
which is the normalizable mode we wish to consider.
For $l=0$, this reduces to the standard result,
$\Phi \sim \rho^{-4}$ quoted above.
(The second solution,
$\psi_{L, \hat L}(\rho) \sim \rho^{l+ {3 \over 2}}$,
 corresponds to the non-normalizable mode.)
As is well known, from the pure $AdS_5$ (Kaluza-Klein reduced) picture,
 the higher spherical harmonics on the $S^5$ correspond to
 massive fields, with mass for the $l^{th}$ mode given by
 $m_{l}^{2} \equiv l(l+4) $.
  This confirms that
  the asymptotic falloff of a field (generated by a compact source)
  is given solely by the $S^5$
  mode  number $l$, and is independent of $\hl$.

 From \assol\ and \sepvar, 
we are interested in the solution with asymptotic behavior
\eqn\phii{
\Phi(\rho,\Omega_3,\Omega_5) \rightarrow \sum_{L,\hat{L}} {M_{L,\hat{L}}\over
\rho^{l+4}}\,
  Y_{\hat{L}}(\Omega_3) \, Y_L(\Omega_5)
}
The coefficients $M_{L,\hat{L}}$ can be related to integrals of  the source
via \multipole. To use this, we need the exact solution to the source-free 
equation which is one at the origin. This is given by a hypergeometric 
function
\eqn\solnu{
R_{l,\hat{l}}(\rho) = \rho^{\hat{l}} \,
F\({\hat{l}-l \over 2},{\hat{l}+l+4 \over 2},\hat{l}+2;-\rho^2\)
}
Setting $u=\sum_{L,\hat{L}} R_{l,\hat{l}}(\rho) \,
  {Y}^*_{\hat{L}}(\Omega_3) \, {Y}^*_L(\Omega_5)$  in \multipole\
(where $*$ denotes complex conjugation)
and using the asymptotic form of the hypergeometric
function for large $\rho$, we obtain
\eqn\M{
M_{L,\hat{L}} = C_{l,\hat{l}} \int_{\Sigma}
s(\rho, \Omega_3, \Omega_5) \, R_{l,\hat{l}}(\rho) \, 
{Y}^*_{\hat{L}}(\Omega_3) \, {Y}^*_L(\Omega_5)  \, f  d^9 V}
where $ f d^9 V = \rho^3 \, d\rho \, d\Omega_3 \, d\Omega_5$,
$\Sigma$ denotes a constant-$t$ slice of \AfSf, and
\eqn\C{
C_{l,\hat{l}} \equiv {k \over 2l+4} \,
{\Gamma({\hat{l}+l+4 \over 2})^2 \over \Gamma(\hat{l}+2) \, \Gamma(l+2)}}

We are interested in the coefficients $M_{L,\hat L}$
since they are directly determined by the
CFT expectation values. More precisely, for each $L$, the asymptotic fields
$\sum_{\hat L} M_{L,\hat L} Y_{\hat L}(\Omega_3)$
are in a one to one correspondence with the expectation value of 
local operators in the gauge theory:
\eqn\keyeq{ \< \O (\Omega_3) Z_L (\Omega_3) \> 
  \propto \sum_{\hat L} M_{L,\hat L} Y_{\hat L}(\Omega_3)}
The operators on the left are defined as follows.
The label $L$ specifies
a particular spherical harmonic on $S^5$ which can be
characterized by a symmetric
traceless tensor $V_{j_1\cdots j_l}$. The CFT operator $Z_L$ is
 $Z_L =V_{j_1\cdots j_l}X^{j_1}
\cdots X^{j_l}$ where $X^j$ are the six scalars in the CFT.
The operator $\O$  depends on which supergravity field in the bulk one is
considering. If $\Phi$ represents the dilaton, then $\O = F^2$.
The total scaling dimension of $\O Z_L$ is thus $l+4$.


How much information about the source is contained in the coefficients
$M_{L,\hat L}$? We have already seen that for $L=0$, corresponding to 
a massless field in $AdS_5$, the coefficients $M_{0,\hat L}$ give all
the multipole moments of an effective source in the $AdS_5$ directions obtained
by averaging the source over $S^5$. For $L \ne 0$, the  coefficients are
related to integrals of the source weighted by a spherical harmonic in the
$S^5$ directions, and a solution to the massive wave equation in the $AdS_5$
directions. As long as the size of the source is much less than the mass of
the field, i.e., small $l$, $R_{l,\hat l}(\rho) \approx \rho^{\hat l}$,
and the coefficients again give standard multipole moments of an
effective source in the $AdS_5$ directions.

We mentioned earlier that if one could
obtain all the coefficients in the
expansion of the source in modes on $S^5$, one could recover the entire
source function in these directions. To see if this is possible,
consider a source which is a product of
a function on $AdS_5$ and a function on $S^5$,  $s= s_1(\rho,\Omega_3) 
s_2(\Omega_5)$. It is clear from \M\ that $M_{L,\hat L}$ is directly
proportional to the coefficient $\sigma_{l,\{m_i\}}$
in the expansion of $s_2$ on $S^5$, and
the remaining integral depends on $l$ but is independent of $\{m_i\}$.
Thus by taking ratios of $M_{L,\hat L}$ with the same $l$ but different
$\{m_i\}$, one can recover $\sigma_{l,\{m_i\}}$ up to one unknown constant
for each $l$ which can be taken to be $\sigma_{l,\{m_i=0\}}$. 
Roughly speaking, one can determine the non-$SO(5)$ invariant part of the
source on $S^5$ directly from ratios of $M_{L,\hat L}$, but one cannot
determine the entire source function.

However, even for $SO(5)$ invariant sources, one can recover a considerable
amount of information. The most interesting case is spherically symmetric
sources in ten dimensions. In this case, the only nonzero $M_{L,\hat L}$
have $\hat L =0$ and $m_i=0$, so they are labeled just by $l$.
The behavior of these coefficients for increasing $l$ is governed by
a competition between the oscillating spherical harmonic on $S^5$, and the
radial function $R_{l,0}(\rho)$, which grows {\it exponentially} with $l$
for $\rho<1$.
This exponential growth is just a consequence of the fact that for small 
$\rho$, $R_{l,0}$ is a solution to a massive wave equation in  nearly 
flat space 
with $m^2_l = l(l+4)$. The solution which is finite at the origin grows
like $e^{m_l \rho}/\rho^{3/2}$ for $\rho \gg 1/m_l$. 
This exponential growth can lead to exponentially
large coefficients. Let us illustrate this with a simple flat 
space example.

Consider four dimensional Minkowski spacetime, ${\cal M}^4$. A conventional
massive field
satisfies \Poiss\ with $k= 4\pi$. Static solutions
behave asymptotically like $\Phi = A e^{-mr}/r$ for some constant
$A$. Suppose the source $s$ is a uniform density ball of radius $r_0$, i.e.,
$s = \Theta(r_0 - r)/{4\over 3} \pi r_0^3$.
Then we can determine $A$ by using \multipole.
The solution to $(\p_i \p^i -m^2) u =0$ with $u=1$ at the origin is
$u=\sinh{mr}/mr$. Substituting into \multipole\ yields
\eqn\expcoef{ A = {3\over  (mr_0)^3}[ mr_0 \cosh{mr_0} -\sinh{mr_0}]  }
So for $mr_0 \gg 1$, $ A \sim  e^{mr_0}/(mr_0)^2$. Of course, outside
the source, this exponentially large coefficient is more than compensated for
by the exponentially small radial dependence, and the solution $\Phi$ is
very small. Similar behavior occurs for small sources in AdS.
But the point is that in the AdS/CFT correspondence, the
radial dependence of the solution is scaled out, and the field theory 
expectation values are directly related to the (exponentially large)
coefficients.

In some cases, this exponentially large contribution to the coefficient
can be canceled by an oscillating contribution coming from compact 
extra dimensions. As a simple example,
consider a massless field in $S^1\times {\cal M}^4$.
A spherical source in this space can be viewed as a periodic
array of sources in ${\cal M}^5$. Since the spacetime is flat, we
know that the asymptotic field can depend only on the monopole moment of 
the source (and the radius of the $S^1$). In particular, it is independent
of the size of the source.
On the other hand, if we expand both the source and the field in
a Fourier series on $S^1$, the coefficient of the asymptotic field
associated with each mode $l$ is given by
an expression similar to \M\ involving the integral of exponentially growing and
oscillating functions. Nevertheless, the integral is independent of the
size of the source for each $l$. In the next section we will see that
this precise cancellation does not extend to curved spacetimes such as
$AdS_5\times S^5$.

\newsec{Size of spherical sources}

We will now turn to one of the most interesting and important implications
of our results, namely, that we can determine the size $r_0$ of a spherical
object.   (By ``spherical'' we mean invariant under SO(9) rotations.)
First, we re-cap the formalism for a more general case of ``bi-spherical'' 
source, defined by $s(\rho, \Omega_3, \Omega_5) = s(\rho, \chi)$ where
$\chi$ corresponds to the radial coordinate on the 5-sphere:
$d\Omega_5^2 = d\chi^2 + \sin^2 \chi \, d\Omega_4^2$.  (Hence, a bi-spherical
source generates a field which is spherically symmetric both on the $S^3$ 
of $AdS_5$ and on the $S^4$ of $S^5$,
i.e.\ $SO(9)$ is broken to $SO(4) \times SO(5)$.)

%
For a bi-spherically symmetric source, $s= s(\rho,\chi)$, 
only the modes with $\hat{l}=0$ and $\{m_i = 0 \}$  will be nontrivial,
 as pointed out in section 2.
That means that we may consider just the spherical
harmonics on $S^5$ which are functions of $\chi$ only.
Eqs. \M\ and \keyeq\ then simplify to 
\eqn\OZdef{
\< \O \, Z_l \> \sim M_{l,\hat{0}} 
\sim \int_0^{\pi}  \int_0^{\infty} s(\rho,\chi) \, R_{l}(\rho) 
\, Z_l(\chi) \, \rho^3 \, \sin^4 \chi \, d\rho \, d\chi
}
where  $R_{l}(\rho) \equiv R_{l,\hat{0}}(\rho)$ 
is given by \solnu\ with $\hat{l} = 0$, and
$Z_l(\chi)$ is proportional to  the 
$\{m_i=0\}$ (real) spherical harmonic $Y_l(\chi)$ 
on $S^5$,  which satisfies the equation\foot{There is a slight abuse of 
notation here. We have previously defined $Z_l$ to be the gauge theory
operator $V_{j_1\cdots j_l} X^{j_1} \cdots X^{j_l}$ where $V$ is symmetric and
traceless. The spherical harmonic $Z_l(\chi)$ denotes (up to a normalization
constant) the function on $S^5$
obtained by taking the same tensor $V$ and contracting all indices with a 
unit vector in $R^6$.}
\eqn\Yl{
Y_l''(\chi) + 4 {\cos \chi \over \sin \chi} \, Y_l'(\chi)
+ l(l+4) \, Y_l(\chi) = 0}
The corresponding solution, expressed in terms of 
hypergeometric functions, is
$Y_l(\chi) = \gamma_l \,
F \( -{l \over 2}, {l \over 2} + 2, {1 \over 2}; \cos^2 \chi \)$ 
for even $l$, and 
$Y_l(\chi) = \gamma_l \, \cos \chi \, 
F \( {1-l \over 2}, {5+l \over 2}, {3 \over 2}; 
\cos^2 \chi \)$ for odd $l$, 
where $\gamma_l$ is a constant fixed by orthonormality of the $Y_l$'s.

Let us now make a few comments on normalization,
implicit in the ``$\sim$'' sign of \OZdef.
First, linearity of the Poisson equation ensures that the 
$\< \O \, Z_l \>$'s of a given source will be directly proportional to the
source's mass $M$.  Thus, for simplicity, we can confine our discussion
to unit mass sources, or equivalently, consider the expectation values
with the mass scaled out. 
Second, since  $\< \O \, Z_l \>$ have different scaling dimensions for 
different $l$, we cannot compare them directly;  we can only meaningfully 
compare operators with the same dimension.  This suggests dividing
 $\< \O \, Z_l \>$ of the given source 
by  $\< \O \, Z_l \>$ of a convenient ``standard reference source'' such as 
unit mass $\delta$-function source.
Thus, we define the normalized expectation values
\eqn\OZnorm{
\OZn \equiv {1 \over M} \, 
{ \< \O \, Z_l \>_s \over  \< \O \, Z_l \>_{\delta}}}
We can treat $\OZn$ as pure numbers
(which depend on the mode $l$ and the source function $s$).
As a by-product, the $l$-dependent coefficients
$C_{l,\hat{0}}$ \C\ coming into the definition \M\ of $M_{l,\hat{0}}$,
as well as the normalization $\gamma_l$ of the spherical harmonics, cancel out.

Evaluating $\< \O \, Z_l \>$ explicitly for a unit-mass $\delta$-function
source, $s_\delta(\rho,\chi) = {\delta(\rho) \over 2 \pi^2 \rho^3} \, 
{\delta(\chi) \over (8\pi^2/3) \,  \sin^4 \chi}$,
then leads to  the following formula:
\eqn\OZ{
\OZn =  {1 \over M}  {16 \pi^4 \over 3} \,
\int_0^{\pi}  \int_0^{\infty} s(\rho,\chi) \, 
F(-{l \over 2},{l \over 2}+2,2;-\rho^2) \, 
{F(-{l \over 2},{l \over 2}+2,{1 \over 2};\cos^2\chi) \over
 F(-{l \over 2},{l \over 2}+2,{1 \over 2};1)}
 \, \rho^3 \, \sin^4 \chi \, d\rho \, d\chi
}
for even $l$, and the analogous expression for odd $l$.
 From the form of \OZ\ it is convenient to normalize $Z_l(\chi)$ such that
$Z_l(\chi = 0) \equiv 1 \ \forall l$. In other words 
 \eqn\Znorm{
Z_l(\chi) \equiv F(-{l \over 2},{l \over 2}+2,{1 \over 2};\cos^2\chi) /
F(-{l \over 2},{l \over 2}+2,{1 \over 2};1)}
  for even $l$, and similarly for odd $l$.
As an immediate consequence of \OZ, for $l=0$, 
$\< \O \, Z_0 \>_n  = \< \O \>_n = 1$  for all sources.

We can use \OZ\ to evaluate $\OZn$ numerically, 
in principle for arbitrary $l$.
However, for wide range of $l$'s, we can obtain the answer much more easily
and efficiently by using series  approximations, which are
 good for small enough values of $l$.
We first describe this in detail and present several results for small $l$.
The most important result is the fact that for any $l > 0$, we can
extract the size of a uniform density,
spherically symmetric source from $\OZn$. 
We then turn to large $l$, where we consider $\OZn$ for the same sources.
We will find that the corresponding results are even more striking.

\subsec{small $l$}

For a small source (of size $r_0 \ll 1$ in AdS units) 
localized on the north pole of $S^5$, 
the integration in \OZ\ only runs over
$\rho \ll 1$ and $\chi \ll 1$, so we can approximate the functions
appearing in \OZ\ by Taylor series in $\rho$ and $\chi$.
In particular, 
\eqn\Rapprox{
R_{l}(\rho) = F(-{l \over 2},{l \over 2}+2,2;-\rho^2)
\approx 1 + {l(l+4) \over 8} \rho^2 
+ {(l-2)l(l+4)(l+6) \over 192} \rho^4 + \cdots
}
and 
\eqn\Zapprox{
Z_{l}(\chi) 
\approx 1 - {l(l+4) \over 10} \chi^2 
+  {l(l+4)(3l^2+12l-8) \over 840} \chi^4 + \cdots
}
(for both even and odd $l$).
For a source with maximal extent  $r_0$, 
$\rho \le r_0 \ll 1$ and $\chi \le r_0 \ll 1$, so that the above series
expansion is a good approximation provided that $l \, r_0 \ll 1$, 
i.e.\ $l \ll {1 \over r_0}$.  The final result, $\OZn$,
will then be expressed as an expansion in powers of $r_0$.

A uniform spherical source of mass $M$ and radius $r_0$
in ten dimensional  flat spacetime would be described by
$ 
s(\rho, \chi) = {M \over V_9 r_0^9}
\, \Theta( r_0^2 - (\rho^2 + \chi^2))
$ 
where $\Theta$ is the Heavyside function and 
$V_9 = {2^5 \pi^4 \over 9 !!}$ 
denotes the volume of a unit ball in $R^9$.
(The denominator normalizes the mass, so that 
$\int s \, d^9V = 
{16 \pi^4 \over 3} \, \int_0^{\pi}  \int_0^{\infty} s(\rho,\chi) \, \rho^3 \, 
\chi^4 \, d\rho \, d\chi 
= M$, independently of $r_0$.)
In exactly flat space, we of course would not  be able to extract the 
size $r_0$ from the asymptotic value of the field.
We will now see how the situation differs in \AfSf.


The \AfSf\ spacetime deviates from the flat ${\cal M}^{10}$ spacetime at 
quadratic order in $\rho$, so we might expect quadratic modifications
to the ``natural'' source.  In fact, there are two distinct modifications.
First, since the mass $M$, given by the integral of
the source, will depend on the measure:
$M = {16 \pi^4 \over 3} \,
\int_0^{\pi}  \int_0^{\infty} s(\rho,\chi) \, \rho^3 \, 
\sin^4 \chi \, d\rho \, d\chi$, the volume normalization $V_9$ will 
receive $r_0$-dependent corrections, $V_9 \to W_9 = V_9 (1 + O(r_0^2))$.
Second, a spherical object should have the same
extent in the AdS and the sphere directions 
in terms of the proper distance, rather than the coordinate distance.  
Specifically, the proper distance in the AdS and the sphere directions
is given respectively by
\eqn\propdist{
\hat{\rho} = \int^\rho {d\tilde \rho \over \sqrt{\tilde \rho^2 +1}} 
  = \sinh^{-1} \rho,
\ \ \ \ \ \ \ \  \hat{\chi} = \chi
}
Hence, a truly spherically symmetric (uniform density) source 
should be written as a function of 
$r^2 \equiv \hat{\rho}^2 + \hat{\chi}^2$:
$$ s(r) 
= {M \over W_9 r_0^9} \, \Theta (r_0 - r) 
= {M \over W_9 r_0^9} \,  \Theta\(r_0^2 - (\sinh^{-1} \rho)^2 - \chi^2\)$$
\eqn\ssource{
\approx {M \over V_9 r_0^9 
(1 + {2 \over 33} r_0^2 + {49 \over 2145}  r_0^4 + \cdots)} \, \,
\Theta\(r_0^2 - \rho^2 (1-{1 \over 3}\rho^2 + \cdots) - \chi^2\)}
which will give $\OZn$ to $O(r_0^6)$. 
(The latter modification will only become relevant at the quartic order.)

\ifig\sls{$\OZn$ from \OZ\ (solid curve), 
and the quartic approximation (dashed curve), for 
uniform density, spherically symmetric source with radius $r_0 = 0.1$.}
{\epsfxsize=8.5cm \epsfysize=5.5cm \epsfbox{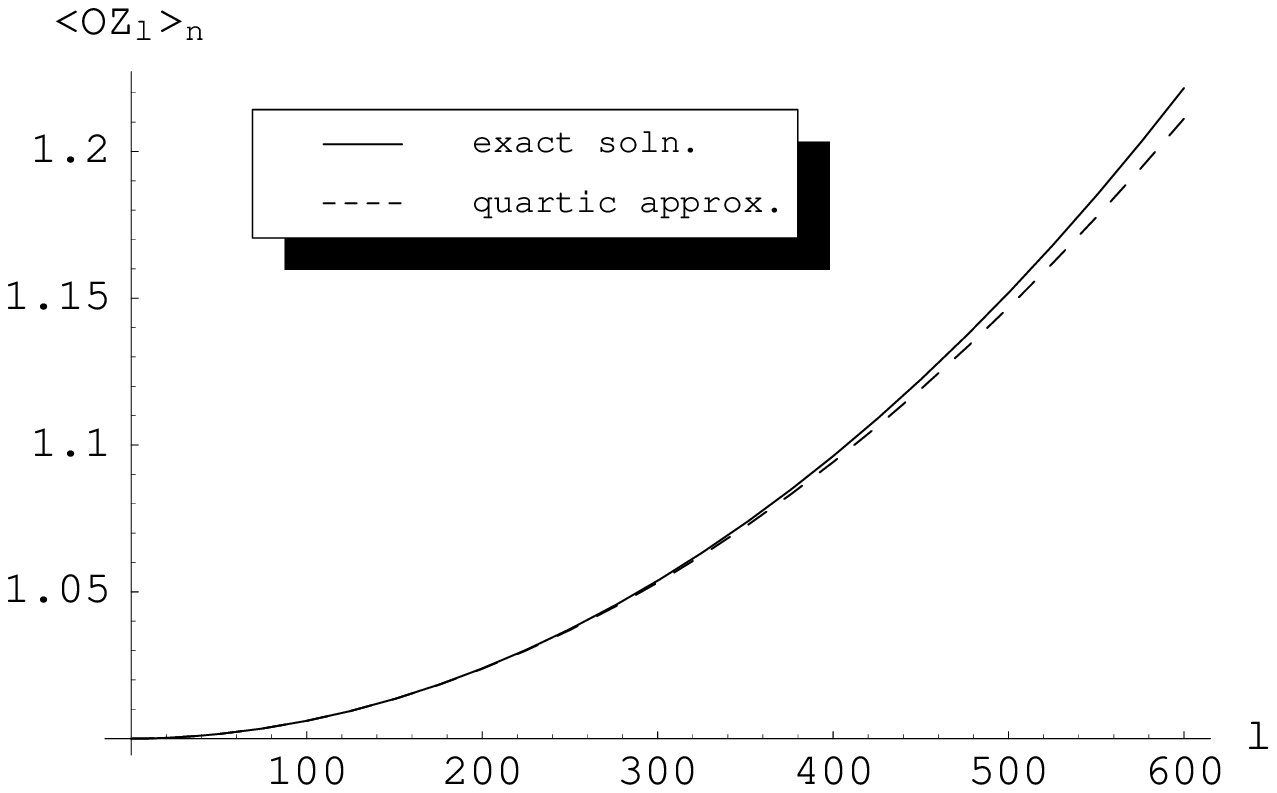}}
We now evaluate $\OZn$ explicitly, up to $O(r_0^6)$, 
for general $l$ (which is small enough so that the series 
expansion in $r_0$ is still useful).
Using the approximations \ssource, \Rapprox, and \Zapprox, we find that 
\eqn\OZunif{
\OZn =  1 + {5 l (l+4) \over 858}  r_0^4 + O(r_0^6)}
Explicitly, if $X_1$ corresponds to the direction of the north pole ($\chi=0$)
on $S^5$, the first few expectation values are:
$$ \< \O \, X_1 \>_n =  1 + {25 \over 858}  r_0^4 + O(r_0^6)$$
\eqn\OZunifEx{
\< \O \, \({6 \over 5} X_1^2 - {1 \over 5} \sum_i X_i X^i \)\>_n =
 1 + {10 \over 143}  r_0^4 + O(r_0^6)}
and so on.

 From \OZunif, we discover that
 the size dependence in $\OZn$ first appears at the quartic order in $r_0$,
and the effect grows only quadratically with $l$.
This is illustrated in \sls, where $\OZn$ is plotted as a function
of the mode number $l$, for a particular small value of 
the source radius ($r_0 =0.1$).
The dashed curve corresponds to the quartic approximation \OZunif,
while the solid curve shows the true behavior.
Before discussing the size dependence itself, let us note two 
interesting facts about the approximation \OZunif.
As is apparent from \sls, the quartic approximation
is valid for a much larger range of $l$ than one would naively expect,
i.e.\ even for $l \gg {1 \over r_0}$ ($ = 10$ in this case).  
In fact, for general $r_0$, the quartic order approximation is good to 
within a fraction of a percent all the way up to $l \sim O(1/r_0^2)$.
Also, the approximation is smaller than the exact behavior, which
seems to be the case for higher order approximations as well.

As we have seen from \OZunif, we may extract the size of the (uniform
density, spherically symmetric) source only at $O(r_0^4)$.
At the first glance, this may seem somewhat surprising, since one 
might have expected that the quadratic deviations of the metric from
flat spacetime should translate into quadratic effect of the size,
i.e.\ $\OZn = 1 + O(r_0^2)$ rather than $\OZn = 1 + O(r_0^4)$.
We may also gain similar expectations from the pure AdS 
case:  The $l$-dependence of the leading asymptotic fall-off of a 
massive field (with mass $m_l \equiv l(l+4)$ and  spherically
symmetric (in $AdS_5$) source of size $\rho_0$   
 is proportional to 
$ 1 + {l(l+4) \over 12} \rho_0^2 + 0(\rho_0^4)$.
The analogous set-up for the full \AfSf\ case would involve a separable
source, such as 
$s(\rho,\chi) \propto \Theta(\rho_0 - \rho) \, \Theta(\chi_0 - \chi)$,
for which we indeed find
\eqn\OZsep{
\OZn =  \(1 + {l (l+4) \over 12}  \rho_0^2  + 0(\rho_0^4)\) \,
\(1 - {l (l+4) \over 14}  \chi_0^2 + 0(\chi_0^4) \)
}
Even when the source is no longer separable, we would still expect
the size information to appear at the 
quadratic level,
 unless there is a very special cancellation.

Such special cancellation at the $O(r_0^2)$ does indeed occur for
spherically symmetric sources.  In fact, for {\it any spherically 
symmetric} source, the $O(r_0^2)$ contribution in $\< O Z_l\>$
will vanish.  
Although we will show this explicitly momentarily, it is perhaps
more instructive to first consider an a-spherical (but still bi-spherical) 
source.  
The simplest such case is the uniform density ellipsoidal source:
$s(\rho, \chi) = {M \over W_9 r_0^9}
\, \Theta( r_0^2 - ({\hat{\rho}^2 \over a^2} + {\hat{\chi}^2 \over b^2}))$,
where 
$W_9 = V_9 \, a^4 b^5 \, 
\(1 + {2 \over 33} (6 a^2 - 5 b^2)  r_0^2  + 0(r_0^4)\)$,
and $\hat{\rho}$,  $\hat{\chi}$ are given by \propdist.
Then we obtain the following result:
\eqn\OZellips{
\OZn =  1 + {l (l+4) \over 22} (a^2 - b^2) r_0^2 + O(r_0^4)}
Thus we see that indeed, for the spherically symmetric case, 
$a=b$, the $O(r_0^2)$ contribution vanishes.\foot{From the quartic term,
which has the $r_0^4$ coefficient 
$${l (l+4) \over 37752} 
\[ 33 \, (a^2 - b^2)^2 \, l^2  + 132 \, (a^2 - b^2)^2 \, l+
4 \, (9 a^4 + 44 a^2  b^2 + 2 b^4) \],$$ we see that the 
$O(r_0^4)$ terms which grow faster than quadratically with $l$ also 
vanish for $a=b$.}
\ifig\sle{$\OZn$ from \OZ\ (solid curve), 
and the quadratic approximation \OZellips\ (dashed curve), for $r_0 = 0.1$,
for an ellipsoidal, uniform density source extended to $r_0$ in
the AdS directions and to (a) ${r_0 \over 2}$ and  (b) $2 r_0$ 
in the sphere directions.}
{\epsfxsize=13.5cm \epsfysize=4.5cm \epsfbox{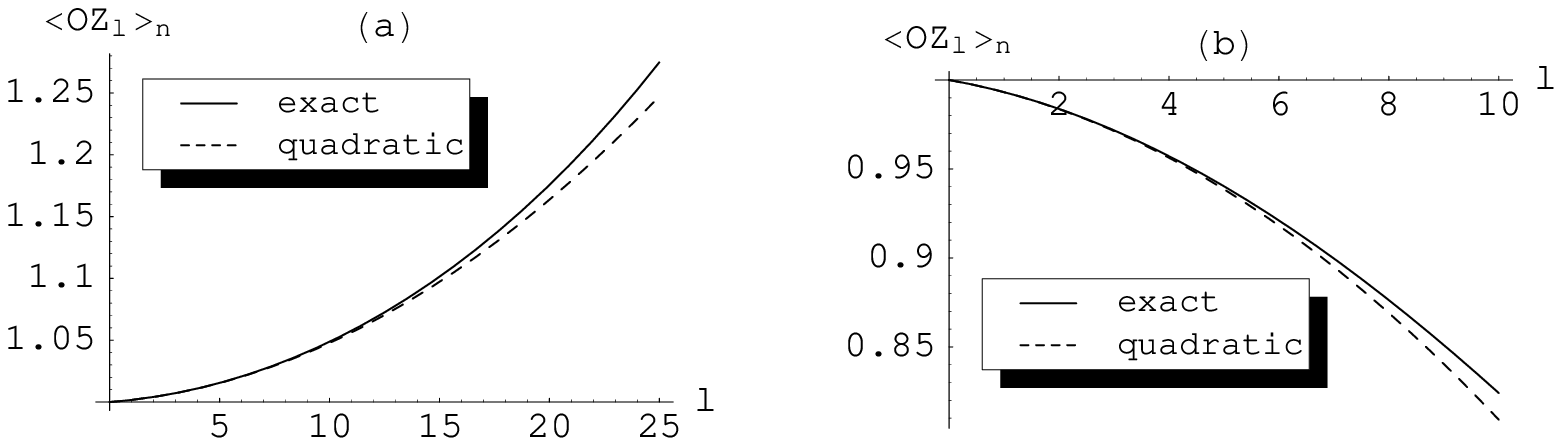}}
On the other hand, for any non-spherical ($a \ne b$) uniform density source, 
the size is visible already at the quadratic level, 
from $(a^2 - b^2) \, r_0^2$.
We show this in \sle, where we plot $\OZn$ (again as a function of $l$)
for $r_0 = 0.1$,
for source extended to $r_0$ in the AdS direction 
and (a) squashed to $r_0/2$ along the  sphere ($a=1, b={1 \over 2}$) 
and (b) stretched to  $2 r_0$ along the  sphere ($a=1, b=2$).
Comparison with \sls\ shows that now we can see the size effect much
earlier in $l$; 
in fact, the relevant scale in $l$ seems to be given by $1/r_0$, 
as we had originally expected,
rather than by $1/r_0^2$, as we discovered for the spherically symmetric
case, $a=b$.

\sle\ also demonstrates another important characteristic of the 
$\OZn$'s for a-spherical sources:  they grow if the source is
more extended in the AdS directions, whereas they decrease if 
the source has a bigger extent along the sphere.  
This is consistent with our naive expectations that 
in the former case, the exponentially growing radial solution in AdS
``wins out'' over the oscillatory contribution from the spherical
harmonic, whereas in the latter case this is reversed: the oscillations
damp out the growing mode.
For the spherical case, then, these two effects balance out
(though unlike in the flat $S^1 \times {\cal M}^4$ example of section 2, 
in \AfSf, they do not cancel completely).

One might wonder whether this special cancellation for spherically symmetric
sources ($a=b$)
had anything to do 
with the source density profile.  We will now show that quite generally,
 for {\it any} spherically symmetric source, the quadratic terms will
cancel.  Although this is intuitively obvious by considering appropriate
superposition of uniform density sources of various sizes, we can see
a simple proof by changing variables to make the spherical symmetry more 
explicit:
\eqn\cv{
\hat{\rho} = r \, \sin \theta, \ \ \ \  \hat{\chi} = r \cos \theta}
where $0 \le \theta \le \pi/2$. 
The source will now be a function of $r$ only, i.e.\ $s = s(r)$,
so that the integration over $\theta$ is independent of the source.
Then, to quadratic order,
\eqn\OZr{
\OZn \propto \int_0^{\infty} s(r) \, r^8 \,
\int_0^{\pi/2} \sin^3\theta \, \cos^4\theta \, 
\( 1 + h_l(\theta) \, r^2 + O(r^4) \) \, d\theta \, dr }
where $h_l(\theta) \equiv - {3 \over 2} \cos^2\theta 
- {l(l+4) \over 10} \cos^2\theta  +  {l(l+4) \over 8} \sin^2\theta $.
The three terms in $h_l$ arise from the integration measure on the 
5-sphere, the spherical harmonic $Z_l(\chi)$ expansion, and the 
radial solution  $R_l(\rho)$ expansion, respectively.)
Since the $l$-dependent part of
$\int_0^{\pi/2} \sin^3\theta \, \cos^4\theta \, h_l(\theta) \,
d\theta$ vanishes (and the $l$-independent part cancels the mass normalization),
the quadratic term disappears, leaving us with
$\OZn \propto \int_0^{\infty} s(r) \, r^8 \, \( 1 + O(r^4) \) \, dr $.
For a source with an extent $r_0$, this yields
$\OZn = 1 + O(r_0^4)$.
One might then worry that there will be even higher order cancellations;
however, performing the above calculation to the next order yields
a nonzero $r_0^4$ coefficient,  consistent
with our previous calculations.

Thus, we have seen that with sufficient precision, we can extract the
size of the object from the $\OZn$ values.  (The fact that for
spherical objects, this effect comes only at the quartic, rather than
quadratic, order, just means that we would need greater precision.)
In all cases, however, the magnitude of the 
size-dependent perturbation of the normalized expectation values
 $\OZn$ increases with increasing $l$.
Before considering the large $l$ regime, 
where this effect is quite pronounced,
let us continue examining how much information about the size can one
obtain from $\OZn$ in the small $l$ regime.

For uniform density ``spherical'' sources, we have demonstrated that
we can extract the size from the $\OZn$ values.
However, extracting the size in this way requires a knowledge of the
source's density profile, apart from spherical symmetry.
In general, we may not wish to rely on such detailed information
(since even for a small star in \AfSf, we don't know its density profile), 
and obtaining drastically different coefficients of the
$r_0$ powers for different profiles would undermine our method's 
usefulness.

Hence, to see if this in fact happens, we now consider 
a spherical source with several different density profiles;
in particular,
\eqn\ssourceq{
s(r) \propto (r_0^2 - r^2)^{\nu} \, \Theta(r_0 - r)}
 where we choose the exponent $\nu = 0, 1, 2,$ and $4$.  
\ifig\slp{For the source density profiles \ssourceq\ with $\nu=0$ 
(solid curves), $\nu=1$
(shortest-dash curves), $\nu=2$ (longer-dash curves),
and $\nu=4$ (longest-dash curves),
(a) the density profiles,
(b) $\OZn$'s from \OZ\ for $r_0 = 0.1$,
and (c)  $\OZn$'s from \OZ\ for 
$r_0 = 0.1$, $0.108$, $0.116$, and $0.129$, respectively.}
{\epsfxsize=14.5cm \epsfysize=4.5cm \epsfbox{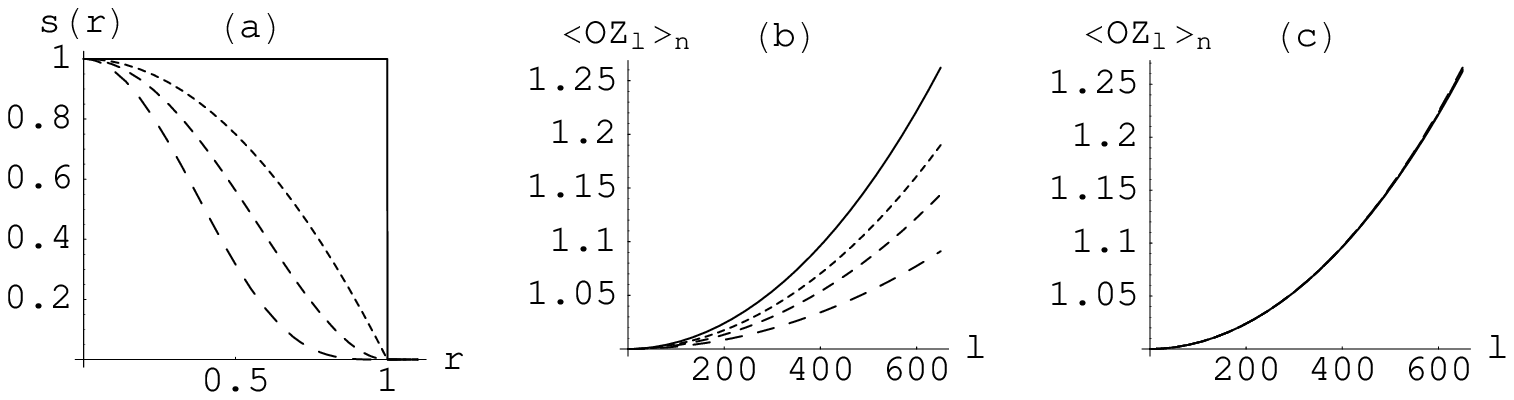}}
\noindent ($\nu=0$ corresponds to the previously 
discussed uniform density case).
These are plotted in \slp a, with $\nu$ increasing
from top curve to bottom curve (i.e.\ the long-dash curve corresponds to the
$\nu = 4$ case).
The corresponding $\OZn$'s (plotted as functions of $l$)  
for $r_0 = 0.1$ are shown in \slp b. 
(The solid curve in \slp b corresponds to the solid curve in \sls.)
The $\OZn$'s behave just as we might expect: the sources with sharper
profiles look effectively smaller, which is reflected in slower
rise of the $\OZn$'s with $l$.

The quartic approximations for the other curves are given by
$$\OZn = 1 + {l (l+4) \over 234}  r_0^4 + 0(r_0^6)
 \ \ \ \ {\rm for} \ \nu = 1$$
$$\OZn = 1 + {l (l+4) \over 306}  r_0^4 + 0(r_0^6)
 \ \ \ \ {\rm for} \ \nu = 2$$
\eqn\OZprof{
\OZn = 1 + {5 l (l+4) \over 2394}  r_0^4 + 0(r_0^6)
 \ \ \ \ {\rm for} \ \nu = 4}
Comparing \OZunif\ with \OZprof,
we find that the sizes $r_0$ required 
to produce the same quartic behavior of $\OZn$ only differ by factors of
$\sim 1.08$, 
$1.16$, and 
$1.29$,  for $\nu=1,2,$ and $4$, respectively.
Thus, if we only had precision up to the quartic order, we could expect to
determine the size of the source (for reasonable density profiles)
only up to factors of order unity.
It turns out that even for the exact solutions for $l < O(1/r_0^2)$, 
the $\OZn$ curves corresponding to different density profiles 
have very similar shapes. 
This is shown in \slp c, where the four $\OZn$'s are 
plotted as in \slp b, except
that the total size of each source is scaled as above, so as to produce
the identical quartic behavior.  Note that these exact curves coincide 
with each other, even 
though they differ from their (common) quartic approximation, as seen in 
\sls.

These results imply that without knowing the source's density profile, 
we could extract its size only up to factors of order unity.
While this would be sufficient for the applications discussed in the
Introduction, we will see that we can actually do better, by
considering the large $l$ regime.

\subsec{large $l$}

We now turn to large $l$.
In particular, we wish to consider the regime where the quartic 
approximations considered above are no longer useful.
One might have expected that this would translate into 
$l \gg {1 \over r_0}$,
since the series expansions of \Rapprox\ and \Zapprox\
 are effectively series in $(l \rho)$ and $(l \chi)$, respectively,
and hence $\OZn$ should be a series in $(l r_0)$.
While this intuition is correct for generic sources
(as confirmed explicitly for the ellipsoidal source cases),
for spherical sources, it turns out that
the special cancellations lead to 
effective series in  $(l r_0^2)$ 
(as was foreshadowed in the previous subsection).
Hence, for the case of spherical sources, ``large $l$'' regime 
will mean $l \gg {1 \over r_0^2}$.\foot{In the AdS/CFT correspondence 
at finite $N$,
there is an upper limit to $l$ coming from the stringy exclusion principle
\mast\ given by $l<N$. However the AdS radius is proportional to $N^{1/4}$
in Planck units, so modes with $l\sim N$ have wavelengths much shorter than
the Planck scale. Since our sources are much larger than the Planck scale,
our large $l$ regime is ${1 \over r_0^2} \ll l \ll N$.}

\ifig\lls{Exponential growth of $\OZn$ at large $l$
 for uniform density spherically symmetric source of size $r_0 = 0.1$:
(a) $\OZn$ and (b) $\ln \OZn$.}
{\epsfxsize=13cm \epsfysize=4.5cm \epsfbox{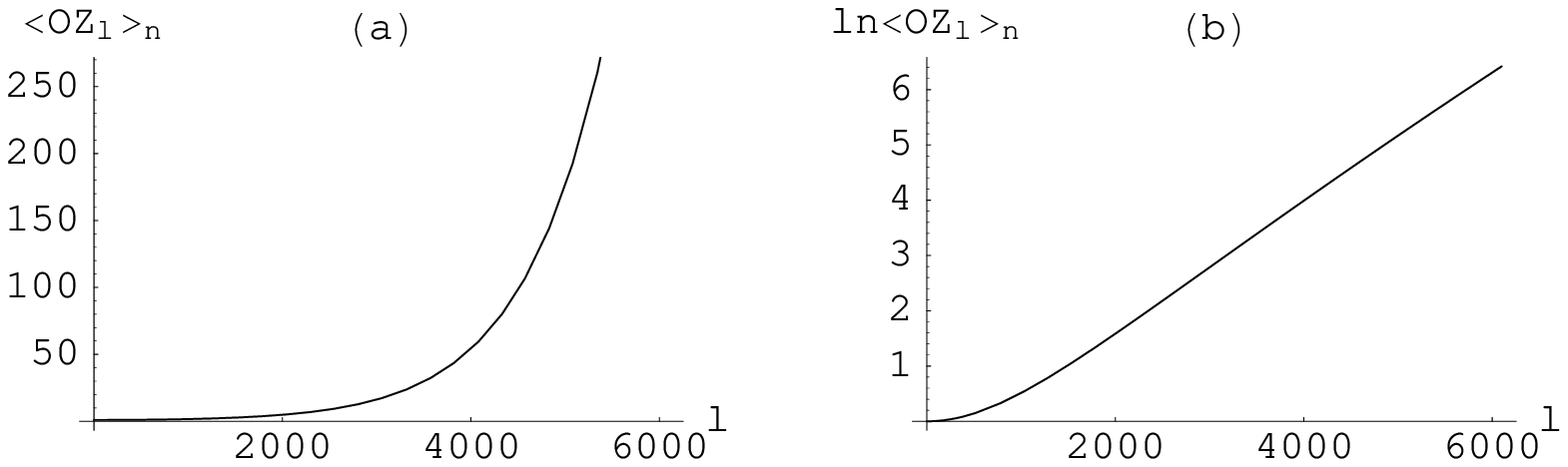}}
We could simply evaluate \OZ\ numerically, 
but it still proves more efficient
to use the series expansion to very high order (such as to $O(r_0^{40})$).
This method is much faster, and we have good control over our accuracy
by examining the relative contribution of the individual terms.
The analogous result as that presented in \sls\ (i.e.\ the continuation
of the solid curve) is shown in \lls a, where we plot $\OZn$ 
for uniform density spherical source of size $r_0 = 0.1$ up to $l \sim 6000$
(where the value of $\OZn$ still has $< 1 \%$ uncertainty).
We now see that $\OZn$ rises much faster than quadratically with $l$; it in 
fact rises exponentially! 
This is verified in \lls b, where the logarithm of $\OZn$ is plotted
as a function of $l$, and asymptotically approaches a straight line.

\ifig\llfs{Exponent $\gamma$ of $\OZn \sim e^{\gamma \, l}$ 
at large $l$
 for uniform density spherically symmetric source of size $r_0$.}
{\epsfxsize=8.5cm \epsfysize=5.5cm \epsfbox{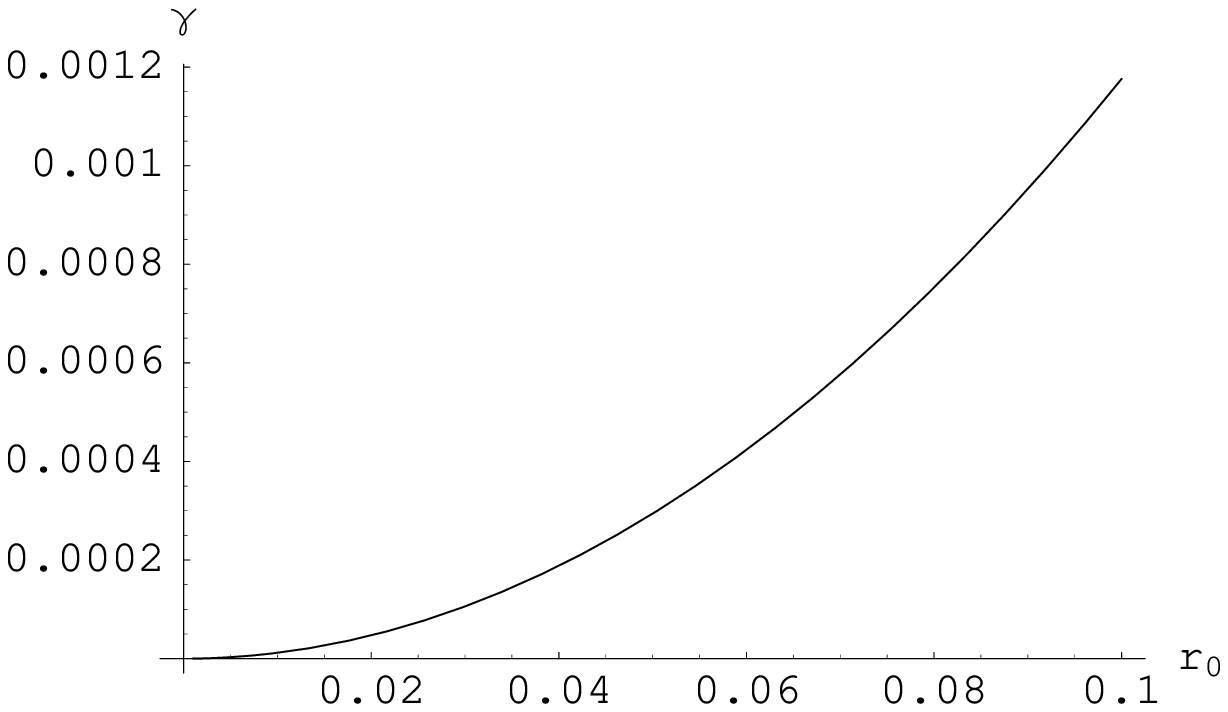}}
The natural question to ask at this point is, how does this exponential
depend on the size $r_0$?
In particular, does $\OZn \sim e^{r_0 \,  l}$ 
as one might expect from the 
$R_l \sim e^{l \rho} /(l \rho)^{3/2}$ exponential behavior,
or does the cancellation seen for small $l$ continue into the large $l$ 
regime?  
The answer is shown in \llfs, where 
(for $\OZn = c \,  e^{\gamma \,  l}$) the exponent $\gamma$ is plotted as 
a function of $r_0$.  Actually, $\gamma$ is approximated very well 
by $\gamma(r_0) = 0.118 \,  r_0^2$, with the two curves indistinguishable
in \llfs.
Numerically, we find that for large $l$,
\eqn\OZasymp{
\OZn \approx 0.49 \, e^{0.118 \, l \, r_0^2}}
Hence, the scaling of $\OZn$ with $r_0$ is indeed slower than one
might have naively expected
(i.e.\ the exponent varies only as $0.118 \, r_0^2$ rather than $r_0$).
This result is nonetheless quite remarkable, since it says that 
in the large $l$ regime, the normalized
\GT\ expectation values are exponentially sensitive to $r_0^2$.

\ifig\llea{Exponential growth of $\OZn$ at large $l$
 for uniform density ellipsoidal source source of size $r_0 = 0.1$
in AdS directions and size ${r_0 \over 2}$ in sphere directions:
(a) $\OZn$ and (b) $\ln \OZn$.}
{\epsfxsize=13.5cm \epsfysize=4.5cm \epsfbox{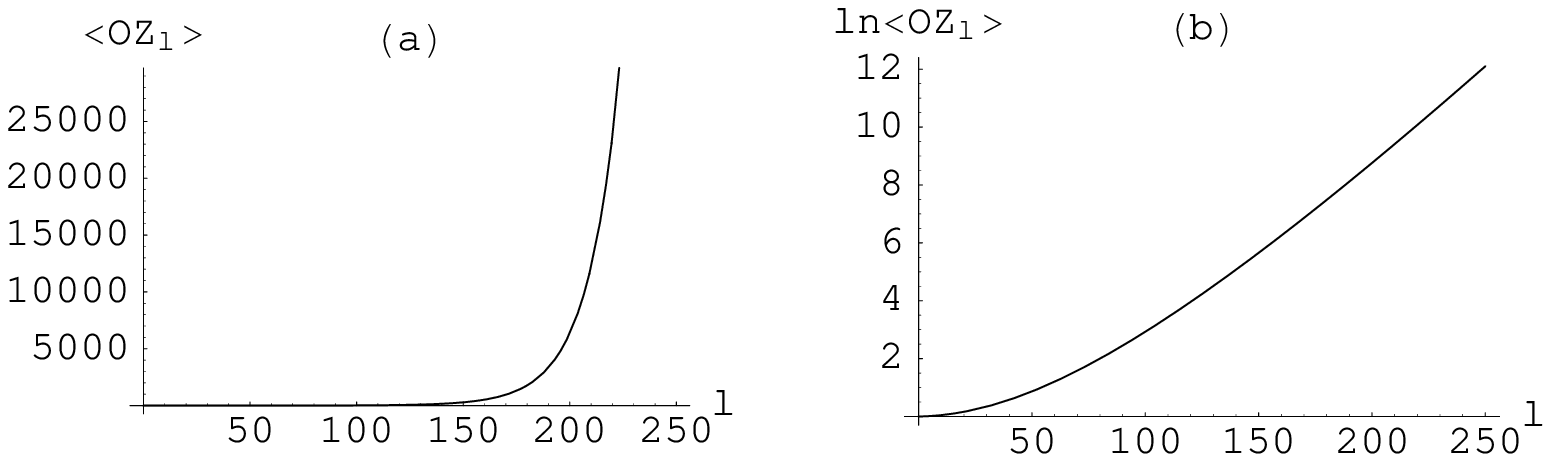}}
The preceding results suggest that  we can obtain even faster growing
$\OZn$'s, if we consider a-spherical sources.  Let us therefore
use the ellipsoidal sources of the previous section (cf.\ \sle).
First, consider the source squashed in the sphere directions, or
equivalently, stretched in the AdS directions.
The corresponding $\OZn$ is plotted (as a function of $l$) in 
\llea a.  From the values of $\OZn$, we confirm that these indeed
grow much faster with $l$ than for the non-squashed case, \lls.
Not too surprisingly,
the behavior again approaches an exponential, 
as seen from the corresponding
plot of the logarithm of $\OZn$, \llea b.

\ifig\llfe{Exponent $\bar{\gamma}$ of $\OZn \sim e^{\bar{\gamma} \, l}$ 
at large $l$
 for uniform density ellipsoidal source of size $r_0$
in AdS directions and size ${r_0 \over 2}$ in sphere directions.}
{\epsfxsize=8.5cm \epsfysize=5.5cm \epsfbox{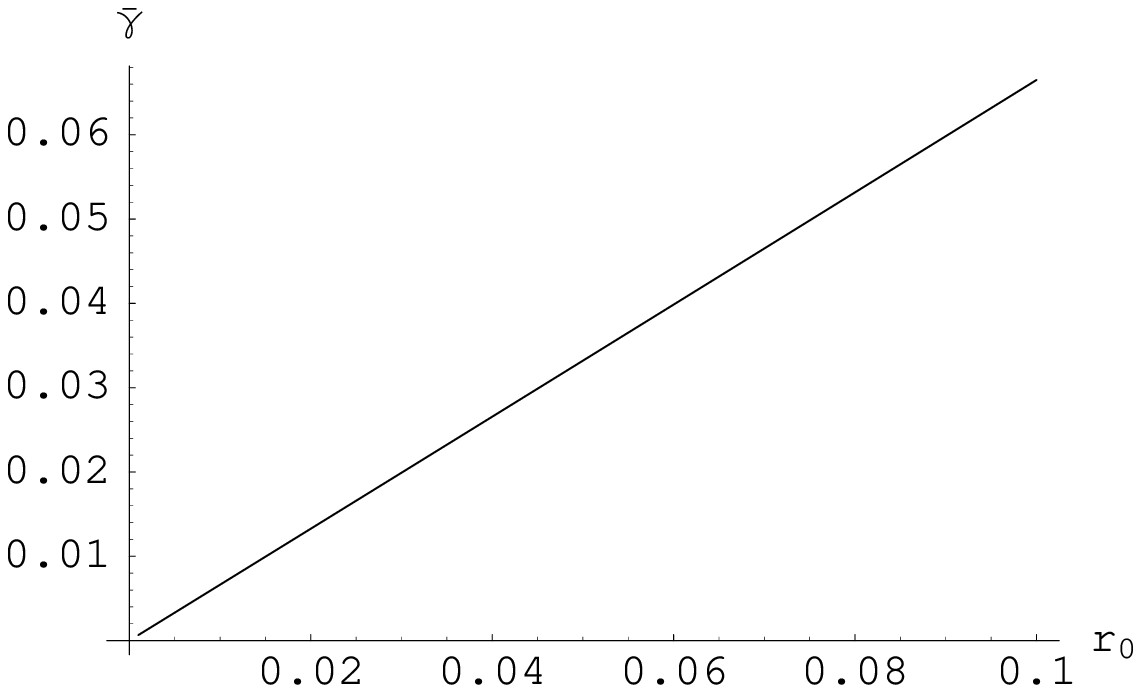}}
As previously, we can analyze the $r_0$ dependence of the
exponent, $\bar{\gamma}$ in $\OZn = \bar{c} \,  e^{\bar{\gamma} \,  l}$.
This is plotted in \llfe\ (which is exactly analogous to \llfs).
Unlike for the spherically symmetric case, the exponent now does
vary linearly with $r_0$, rather than quadratically.
(This gives rise to the faster growth of these expectation values.)
More specifically, we find that the behavior of $\OZn$ for large
$l$ may be approximated by
\eqn\OZelasymp{
\OZn \approx 0.01 \, e^{0.67 \, l \, r_0}}

\ifig\lles{For uniform density ellipsoidal source source of size $r_0$
in AdS directions and size $2 r_0$ in sphere directions:
(a) $\OZn$ from \OZ\  for  $r_0 = 0.1$, 
(b) inverse $l$ at which $\OZn$ first crosses zero for given $r_0$.}
{\epsfxsize=13.5cm \epsfysize=4.5cm \epsfbox{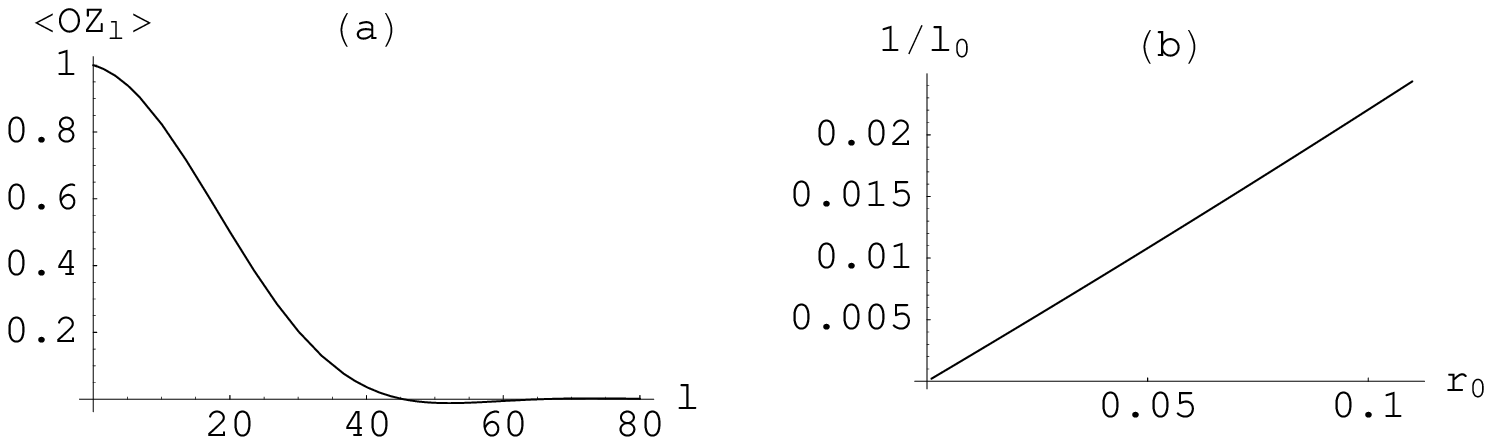}}
Let us now briefly consider the source extended in the sphere
directions.  The continuation of $\OZn$ from \sle b to larger
values of $l$ (namely, to $l \sim 80$) is given in \lles a.
The initial drop of $\OZn$ is followed by damped oscillations
around $\OZn = 0$.  
While this is perhaps not as glamorous as exponentially growing
$\OZn$'s, it is certainly a significant signature of the 
size of the source.  Furthermore, this case illustrates that
$\OZn$ can actually change sign for appropriate $l$.
For completeness, we can extract the characteristic $r_0$ dependence
by considering, for instance, $l \equiv l_0$ 
defined as the $l$ for which $\OZn$ first vanishes.
This should be inversely proportional to $r_0$, since for smaller
sources, $\OZn$ should fall off slower with $l$ and therefore cross
zero later.  Plotting $1/l_0$ for different $r_0$, 
as shown in \lles b, 
indeed verifies this to be the case.

The above results, 
summarized in \llfs, \llfe, and \lles b,
 confirm our expectation that the characteristic 
scale for $l$ of $\OZn$ 
is $O(1/r_0^2)$ for spherically symmetric sources
but only $O(1/r_0)$ for ellipsoidal sources.
Thus, ellipsoidal sources in \AfSf\ (extended in the AdS)
would have an even more spectacular signature in the \GT.
Although we presented the results for rather large a-sphericities
(where the extent of the source in the AdS and sphere directions
differed by a factor of 2), we find that the same basic behavior 
holds for any ellipsoidal source with non-zero a-sphericity.
(This is a direct extension of the behavior summarized in \OZellips.)

We have considered the behavior of $\OZn$ for ellipsoidal sources
mainly to contrast this with the behavior of the (presumably more
physically relevant) spherically symmetric sources.  
However, one can ask, how physically relevant {\it are} the
ellipsoidal sources (perhaps with very tiny a-sphericities)?  
In other words, do we expect, say, ``stars''
in \AfSf\ to be exactly spherically symmetric?
We cannot answer this question without knowing the detailed 
equation of state, etc.  However, we note that since the tidal forces
will be different in the AdS and the sphere directions, 
spherically symmetric density profile of the star would preclude
spherically symmetric pressure profile.  Conversely, for spherically 
symmetric pressure, the source can't be expressed as a function of $r$.
Such a source might then look more like an ellipsoid
(presumably squashed in the $\rho$ direction due to the AdS potential).

\ifig\llp{For the source density profiles \ssourceq\ with $\nu=0$ 
(solid curves), $\nu=1$
(shortest-dash curves), $\nu=2$ (longer-dash curves),
and $\nu=4$ (longest-dash curves),
(a) $\OZn$ and (b) $\ln \OZn$ for $r_0 = 0.1$;
(c)  $\OZn$  and (d) $\ln \OZn$ for 
$r_0 = 0.1$, $0.108$, $0.116$, and $0.129$, respectively.}
{\epsfxsize=13.5cm \epsfysize=8.5cm \epsfbox{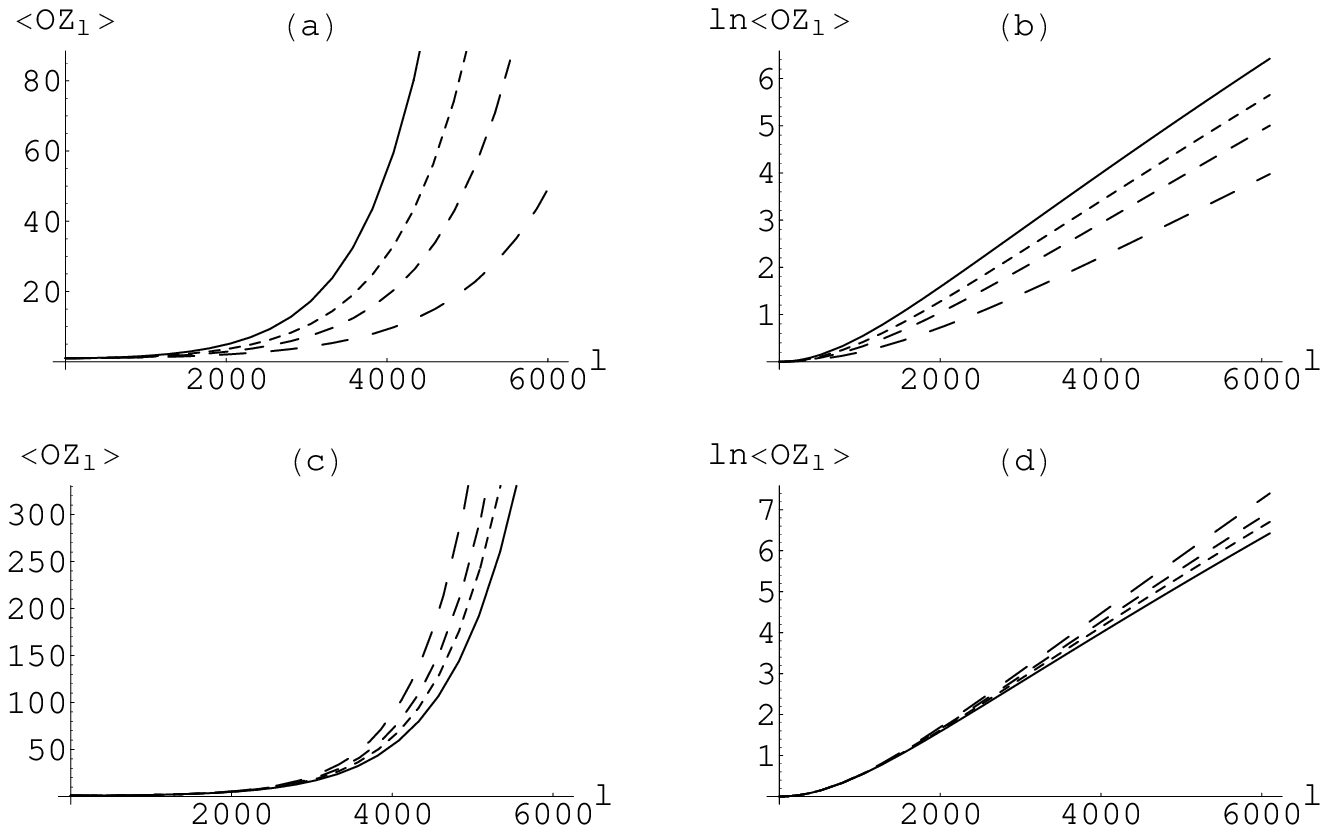}}
Finally, to complete the large $l$ discussion in parallel
with the small $l$ analysis above, let us now consider spherical
sources with various density profiles.
We have seen that in the small $l$ regime,
we could not distinguish certain sources with different
profiles and correspondingly different sizes from each other.
Since at large $l$, $\OZn$ becomes exponentially sensitive
to the size, one might hope that in this regime the expectation
values would allow us to distinguish these various sources.
Such hope is indeed realized, as we show in \llp.
In \llp a and \llp c we plot the extensions of the curves in 
\slp b and \slp c, respectively, up to $l \sim 6000$
(same sources and same conventions as used in \slp\ are used here as well).
 From \llp c, we clearly see that we can distinguish the various
sources for large enough $l$.  
All of these curves exhibit the large $l$ exponential behavior 
characteristic of the spherically symmetric sources, as
shown in \llp b and \llp d, which are just the respective logarithmic plots 
of \llp a and \llp c.

\newsec{Generalization to $AdS_p \times S^q$}

Above, we have encountered a special cancellation of the
leading size signature for a spherical source in \AfSf.
One may wonder whether this is a result of the fact that the curvature
of the AdS and 
sphere have  the same magnitude but opposite sign.
An affirmative answer would imply that our results could be drastically 
different
in other spacetimes of interest for the AdS/CFT correspondence,
namely in \AfSs\ and in \AsSf,
where the radii of curvature are different for the
AdS and the sphere.
To check this, we now extend the previous calculation to the
more general case of $AdS_p \times S^q$.

The spacetime is described by the metric
\eqn\pqmetric{
ds^2  \, = 
- (\rho^2 + 1) \, dt^2 + {d\rho^2 \over \rho^2 + 1} + \rho^2 \, d\Omega_{p-2}^2  
+ \alpha^2 \, (d\chi^2 + \sin^2 \chi \, d\Omega_{q-1}^2)}
As in the \AfSf\ case, we set the radius of curvature of AdS to unity,
so that $\alpha$, which can be written as ${q-1 \over p-1}$,
 gives the radius of the sphere.
The proper distance along the sphere direction now depends on $\alpha$, so
that \propdist\ becomes
\eqn\propdistpq{
\hat{\rho} = \sinh^{-1} \rho, \ \ \ \ \ \ \ \  \hat{\chi} = \alpha \, \chi
}

The solution to the radial equation which is one at the origin is
$$R_{l}(\rho) = 
F(-{\lambda \over 2},{\lambda \over 2}+{p-1 \over 2},{p-1 \over 2};-\rho^2)$$
\eqn\Rpq{
\approx 1 + {\lambda (\lambda +p-1) \over 2 (p-1)} \, \rho^2 
+ {\lambda (\lambda -2) (\lambda +p-1) (\lambda +p+1) \over 8 (p^2-1)} \, \rho^4
+ \cdots}
where $\lambda \equiv {l \over \alpha} = l \( {p-1 \over q-1} \)$.
The spherical harmonic is given by 
$Y_l(\chi) \propto
F \( -{l \over 2}, {l \over 2} + {q-1 \over 2}, {1 \over 2}; \cos^2 \chi \)$ 
for even $l$, and 
$Y_l(\chi) \propto \cos \chi \, F \( {1-l \over 2}, {l+q \over 2}, {3 \over 2}; 
\cos^2 \chi \)$ for odd $l$.
The small-$\chi$ expansion of the spherical harmonic 
(normalized to 1 at the north pole) is
\eqn\Zpq{
Z_l(\chi) \approx 1 - {l (l+q-1) \over 2q} \chi^2 +
{l (l+q-1) \( l^2 + (q-1)l - {2 \over 3} (q-1) \) \over 8 q (q+2)} \chi^4 + \cdots}
Finally, the volume form is modified, so that for a bi-spherical source
on this spacetime, the coefficient of the asymptotic field  
becomes\foot{Although we are
not considering expectation values here (since there is no dilaton in these
cases),
the massless scalar field can be viewed as a toy model for the linearized
supergravity fields.}
\eqn\OZpq{
M_{l,\hat 0} \propto
\int_0^{\pi}  \int_0^{\infty} s(\rho,\chi) \, R_{l}(\rho) 
\, Z_l(\chi) \, \rho^{p-2} \, \sin^{q-1} \chi \, d\rho \, d\chi
}

The small-$l$ behavior of $M_{l,\hat 0}$ 
for a uniform density spherical source 
$s(\rho, \chi) \propto
 \Theta( r_0^2 - (\sinh^{-1} \rho)^2 - \alpha^2 \, \chi^2))$
is then given by 
\eqn\OZunifAf{
M_{l,\hat 0} \propto   1 + {13 l (l+6) \over 16128}  r_0^4 + 0(r_0^6)}
for \AfSs, and
\eqn\OZunifAs{
M_{l,\hat 0} \propto  1 + {5l (l+3) \over 126}  r_0^4 + 0(r_0^6)}
for \AsSf.
This shows that these cases are qualitatively similar to \AfSf.
The size continues to appear at order $r_0^4$
and not at order $r_0^2$.

\newsec{Discussion}

We have seen that the expectation value of local operators in the CFT
contain a considerable amount of information about small sources inside
$AdS_5\times S^5$.
The most striking result is that, for a spherical source of radius $r_0$,
the appropriately normalized expectation values of the operators
$\O Z_l$ (defined after
\keyeq) grow exponentially with $lr_0^2$ \OZasymp. So the 
size of a small source has a big effect on the expectation values. This
exponential growth is obtained at large $l$. For small $l$, the size dependence
first arises in a term of order $r_0^4$ \OZunif. 
As discussed in section 3, one might have expected
the size dependence to be an even larger effect. Since the curvature is
noticable at $O(r^2)$, it could
first show up at order $r_0^2$ for small $l$, and
grow like the exponential of $l r_0$ for large $l$. It is not
yet clear what is special about the spacetimes $AdS_p\times S^q$ arising
in the AdS/CFT correspondence to cause this extra cancellation.  In this
regard, it would be
interesting to consider spacetimes with less supersymmetry, e.g., $AdS \times
K$. While we still expect a small source to produce exponentially growing
expectation values, it is not clear whether the rate will be governed
by $l r_0$ or $l r_0^2$.

We have also found an unusual behavior  of the large $l$ expectation
values for ellipsoidal sources.
When the source is slightly elongated in the AdS directions, the
expectation values again grow exponentially, but when the source is slightly
elongated in the sphere directions, they are quickly damped to  
zero (Fig. 8). While there are many examples of different supergravity 
configurations with the same local expectation values e.g. \refs{\dkk,\giro}, 
it is surprising that very
similar supergravity configurations can yield vastly different
expectation values.

We have considered static sources at the center of $AdS_5$. The
extension to time dependent sources remains to be investigated, and is likely
to have interesting consequences. For example, consider a source which is
slowly oscillating from being extended in the $AdS_5$ directions to being
slightly extended in the $S^5$ directions. It would appear that
the normalized expectation values of operators $\O Z_l$
with large $l$ in the field theory must change rapidly
from being exponentially large
to almost zero  in each oscillation. As a second example, suppose one gives the
center of mass a small velocity. Even though the spacetime is locally
approximately Poincare invariant, the gauge theory description depends
crucially on whether the velocity is in the $S^5$ or $AdS_5$ directions.
If it is in the $S^5$ direction, the scalars in the nonzero expectation values
will change, but the dependence on the $S^3$ at infinity will be unaffected.
If the velocity is in the $AdS_5$ direction, the source will oscillate
following a geodesic. This will cause the dependence on the $S^3$ to change,
but the scalars in the expectation value will remain unchanged. For the
special case of a point
source, the expectation value  for the $l=0$ mode was computed in \refs{\hoit,
\dkkk}. One
finds that the radial position of the source
is correllated with the size of the excitation in the gauge theory, in accord
with the UV/IR correspondence.

One application of our results is to the description of small black holes.
With $AdS_5\times S^5$ boundary conditions,
it is possible for small ten dimensional black holes to be in static
equilibrium with their Hawking radiation \gary. How would the CFT distinguish
these states from other states with the same total energy? 
A state of pure
radiation with this energy would not be localized on the $S^5$, and so would
have zero expectation values for operators of the form $\< O Z_l\>$. 
Since we do not yet
know the exact supergravity solution corresponding to a small black hole
in $AdS_5\times S^5$, finding the field theory expectation
values in this state remains an open problem. The
results obtained here are not applicable due to nonlinear effects near the
horizon. However, gravity is essentially linear for an object just ten
times larger than a black hole,  and we have seen that this would be reflected
in the expectation values growing exponentially with $l$, with
an exponent that depends on the radius of the star. 
So one could easily rule this out. Once one
knows that there is a rather large mass contained within about ten Schwarzschild
radii, the only possibility is a black hole. Although convincing, this
argument is still indirect. We do not yet have a good description of the
spacetime causal structure directly in the gauge theory.
Note that one signal of
an evaporating black hole is that the bound on the size of the source will
decrease with time.

We remark that an alternate way to discern a black hole in the bulk
through the local boundary operators
would be by the use of probes.  Namely, a suitably-designed probe
thrown into a \BH\ would never reemerge, unlike the same probe 
thrown into a star or a thermal gas of radiation.  Thus, after
a sufficient time, we could learn about the bulk through the observed
behavior of the probe.
However, although this method would yield a more direct detection
of the event horizon, and would still use just the local operators in the \GT,
it requires following these operators for a time $\Delta t \sim \pi R$.

We have seen that asymptotically $AdS_5\times S^5$ boundary conditions make
a holographic description easier than for asymptotically flat spacetimes,
since one can
recover much more information about small objects from  the asymptotic fields.
Whether a purely holographic description exists for 
asymptotically flat spacetimes remains to be seen.

\vskip 1cm

\centerline{\bf Acknowledgements}
It is a pleasure to thank M. Rangamani,
L. Susskind, and W. Taylor for discussions.
This work was supported
in part by NSF Grants PHY-9507065 and PHY-0070895.
%
%
\listrefs
\end